\documentclass[a4paper,11pt]{article}
\usepackage{amsmath,amssymb,mathrsfs}
\usepackage[top=3cm, bottom=3cm, left=3cm, right=3cm]{geometry}
\date{}
\begin{document}
\author{Burak Tevfik Kaynak\\ Feza G\"{u}rsey Institute\\
Kandilli 34684, Istanbul, Turkey\\ kaynak@gursey.gov.tr}
\title{\bf Quantum Effective Action in Odd Dimensions and Zeta-Function Regularization}
\maketitle
\begin{abstract}
In this work, we mainly study the one-loop effective action for real scalar theories in non-homogeneous backgrounds in odd dimensions. It is shown that through the method studied in Ref. \cite{kaynak}, it is possible to obtain a unified result for the one-loop correction to the Euclidean effective action in $2+1$ dimensions. Our result simplifies into the ones given in the literature calculated via different techniques based on some assumptions of field and its derivative. The one-loop correction is obtained in $(2n+1)$-dimensional Euclidean space-times, as well. Moreover, its generic behavior is discussed for non-homogeneous backgrounds. Afterwards, the structure of the divergence for scalar field theories at one-loop is discussed by studying the one-loop correction in $2n$-dimensional Euclidean space-times. The beta function of the $O(N)$-invariant nonlinear $\sigma$-model in $d=2+1$ is calculated at leading order in the $1/N$ expansion without using the $\varepsilon$-expansion method.
\end{abstract}
\section{Introduction}
The notion of quantum effective action plays an indispensable role in relativistic quantum field theories. It can concisely be expressed as the Legendre transform of the generating functional of connected correlation functions, which is analogous to the free energy of statistical physics. Since the quantum effective action generates the full irreducible vertex functions of the theory, it is in general called the generating functional of proper vertices. Especially, its specialization to constant fields, the so-called effective potential, is of unquestionable importance, as pointed out in Ref. \cite{lasino}, such that it allows us to study spontaneous symmetry breaking mechanism, not only in relativistic quantum field theories coupled to gauge fields, which are prominent for particle physics, but in statistical field theory and non-relativistic many body systems, as well \cite{schwinger,dewitt,thooft,coleman,justin}. What is known from the perturbation theory that this generating functional is identical to the classical action at the leading order as far as the loop expansion is concerned. This is another reason why it attracts attention for the discussion of spontaneously broken theories. Moreover, it has a natural non-perturbative generalization as a flow equation in the exact renormalization group program \cite{wetterich}.

From the loop expansion point of view, the one-loop correction to the effective action, $\Gamma^{(1)}$, is basically the generating functional of the one-particle irreducible (one-loop) Feynman diagrams. Unfortunately, it is generally ill-defined since it is given by a functional determinant of elliptic operators. Therefore, one has to regularize it so as to obtain meaningful results. The zeta function regularization is an elegant method of defining such determinants, and it is known that any quantum field theory can miraculously be renormalized via its analytical continuation properties at one-loop \cite{hawking,ball}. The heat kernel is widely used in order to compute the zeta function of the operator under consideration, and the literature on this subject is vast. One of the reasons why it is a generally chosen method stems from the fact that it permits a short-time expansion, whose coefficients can be given in terms of geometric invariants that are of great use for theories involving gauge theories, especially.

In Ref. \cite{kaynak}, an alternative method for calculating a semi-classical expansion for infinite dimensional determinants in the case of non-constant backgrounds was studied, based on the idea of finding an expansion for the arbitrary complex powers of elliptic differential operators in pseudo-differential operator language as in \cite{seeley}. The authors in Ref. \cite{kaynak} show that recognizing not only the leading momentum part, which is formerly regarded as the principal symbol, but also the potential part of the symbol altogether as the principal symbol of the pseudo-differential operator allows us to obtain naturally a general derivative expansion, having desired terms which are of importance in quantum effective actions. This modification actually has a very profound effect on the derivative expansion, in the sense that the derivative expansion calculated via the short-time expansion of the heat kernel is based on constructing an asymptotic expansion whose first term is related to nothing but the kinetic operator of the free theory, $-\partial ^2$. However, in the modified resolvent method, the first term directly corresponds to the operator $-\partial^2 + V(x)$ instead of $-\partial^2$. Hence, each term in the expansion corrects that operator. This approach can also be extended to the case in which the space-time is curved \cite{gusynin1,gusynin2}.

The organization of the paper is as follows: in Sec. \ref{sec2}, following Ref. \cite{kaynak}, the one-loop correction to the effective action for the self-interacting scalar field theory without specifying the form of the interaction in advance will, first,be calculated in $(2+1)$-dimensional Euclidean space-time. It will be shown that the technique previously developed by the authors in the mentioned reference yields a unified answer as a derivative expansion such that it contains the results given in the literature, which are previously calculated via different techniques with predetermined assumptions both on the field and its derivative depending whether the background is non-constant or constant. It will be demonstrated that our result simplifies into the results obtained in the literature previously in the aforementioned limits. Secondly, the one-loop correction will be obtained for generic odd dimensions, and the general features of the results be pointed out. Afterwards, why the result for this case does not have any divergences will be discussed.

In Sec. \ref{sec3}, with the purpose of studying the structure of divergence for  self-interacting scalar theories, the one-loop correction will be calculated and examined in generic even dimensions. The transition between two renormalization prescriptions, namely the zeta-function and the dimensional regularization for the massless $\phi^4_4$ theory, will briefly be illustrated for the sake of completeness.

In Sec. \ref{sec4}, as an application, the $O(N)$-invariant nonlinear $\sigma$-model will be revisited. The beta function of the theory at leading order in the $1/N$ expansion will be calculated through the effective action obtained for the auxiliary field, which is introduced to keep the constraint coming from the $O(N)$ invariance. Even though the construction of the model takes place in $d=2+1$, the beta function will be calculated without the $\varepsilon$-expansion method.  
\section{Quantum Effective Action in Odd Dimensions} \label{sec2}
In this section, we will use the method which was developed in Ref. \cite{kaynak} in order to evaluate the one-loop effective action for bosonic operators in three dimensions first, and in generic odd dimensions afterwards. For the reader's convenience, the essential tools upon which the method is based will be reviewed. We would like to invite the reader to look at Ref. \cite{kaynak} for the details of the calculations, and to look into both Refs. \cite{ball,elizalde} and the ones cited therein for comprehensive guides to the field.

The one-loop correction to the Euclidean effective action in the case of real scalar field can be given by
\begin{align} \label{qeff}
   \Gamma^{(1)} &= \frac{1}{2} \log \det A\,,
\end{align}
where $A$ is typically a second order elliptic differential operator. One has to give a meaning to the right-hand side of Eq.~(\ref{qeff}). One of the well established methods to define such an infinite dimensional determinant is to introduce the zeta function for the differential operator $A$, and then to regularize it by the analytical continuation properties of the zeta function. The one-loop correction to the effective action, then, reads,
\begin{align}\label{deffact}
   \Gamma^{(1)} &= -\frac{1}{2} \zeta' ( 0 | A ) \,,
\end{align}
where $\zeta(s|A)$ is the zeta function associated with the operator $A$, and the prime stands for the derivative with respect to $s$. The zeta function can be given by
\begin{align} \label{zeta1}
   \zeta(s|A) &= \mathrm{Tr} A^{-s} = \int d^d x \zeta(s|A)(x) \,,
\end{align}
$\zeta(s|A)(x)$ being the local zeta function, which is a meromorphic function of $s$ having merely simple poles on the real axis. $\zeta(s|A)$ and its first derivative are regular at $s=0$.

The semi-group integral representation of the inverse complex powers of an operator endowed with the symbol calculus of the pseudo-differential operators is well suited to provide a semi-classical expansion for the one-loop effective action. Since the semi-group representation depends on the resolvent of the operator, and the symbol calculus allows us to construct a very powerful machinery to evaluate the symbol of the resolvent as a semi-classical expansion. Therefore, obtaining an expansion for the resolvent automatically leads to the desired expansion for the effective action.

In that manner, the zeta function can be given by
\begin{align}\label{zeta2}
   \zeta(s|A) &= \frac{\sin \pi s}{\pi} \int d^d x \int \frac{d^d p}{(2 \pi)^d} \int_0^\infty d \lambda \, \lambda^{-s} \sigma \left(\frac{1}{\lambda + A} \right) \,,
\end{align}
$\sigma(A)$ be the symbol of the operator $A$, and the trace in the definition of the zeta function in Eq.~(\ref{zeta1}) transforms into a phase space integral above. If the operator under consideration has discrete indices, then one should also take another trace over these indices, such as Dirac operators. The symbols used in this paper are basically Weyl ordered and treated as matrix valued distributions on the phase space $\mathbb{R}^n \otimes \mathbb{R}^n$. Since it is a well known fact that if the Weyl ordered symbols are used to represent the associated operators on that space, there occurs, then, a one-to-one correspondence between functions on the phase space and the operators acting on the Hilbert space $\mathcal{L}^2(\mathbb{R}^n)$, well described by Ref. \cite{hormander}.

Let $A$ be a pseudo-differential operator given by the formula, \cite{shubin},
\begin{align}
   A u(x) &= \int d^d y \int \frac{d^d p}{(2 \pi)^d} \tilde{A} \left( \frac{x+y}{2},p\right) e^{i p \cdot (x-y) } u(y) \,,
\end{align}
in which $\tilde{A}$ is nothing but the symbol $\sigma(A)$, used for brevity, and defined by the Fourier transform of its kernel with respect to the relative coordinates,
\begin{align}
   \tilde{A}(x,p) &= \int d^d \xi \, A\left(x + \frac{\xi}{2}, x - \frac{\xi}{2}\right) e^{- i p \cdot \xi} \,.
\end{align}

It turns out that the multiplication between the operators introduces a new multiplication rule between their symbols. It can be shown that this induced multiplication, the so-called star product, is non-commutative, associative, and can be given by
\begin{align}
\tilde{A} \circ \tilde{B} &= \exp \left[ \frac{i\hbar}{2} \left( \frac{\partial}{\partial x^\mu} \frac{\partial}{\partial p'_\mu} - \frac{\partial}{\partial p_\mu} \frac{\partial}{\partial x'^\mu}\right) \right] \tilde{A}(x,p) \tilde{B}(x',p') \bigg|_{x=x';p=p'} \,.
\end{align}
This exponential can also be given in term of the generalized Poisson brackets as an asymptotic expansion,
\begin{align}
\tilde{A} \circ \tilde{B} &= \sum^\infty_{n=0} \left( \frac{i\hbar}{2}\right)^n \frac{1}{n!} \left\lbrace \tilde{A} , \tilde{B} \right\rbrace_{(n)} \,,
\end{align}
where the generalized Poisson brackets are defined as
\begin{align}
\left\{ \tilde{A},\tilde{B} \right\}_{n}&= \sum^n_{i=0} ( -1)^n \tilde{A}^{\nu_1 \cdots \nu_i}_{\mu_1 \cdots \mu_{n-i}} \tilde{B}^{\mu_1 \cdots \mu_{n-i}}_{\nu_1 \cdots \nu_i} \, ,
\end{align}
in which $\tilde{A}^{\mu_i} = \frac{\partial \tilde{A}}{\partial p_{\mu_i}}$ and $\tilde{A}_{\mu_i} = \frac{\partial \tilde{A}}{\partial x^{\mu_i}}$.

If one lets the symbol of the resolvent be expanded in power series in $\hbar$ as
\begin{align}
   \sigma\left( \frac{1}{\lambda + A} \right) &= \sum_{n=0}^\infty \hbar^n \tilde{R}_n(\lambda) \,,
\end{align}
then the following condition,
\begin{align}
   \sigma\left( \frac{1}{\lambda + A} \right) \circ \sigma( \lambda + A) &= 1 \,,
\end{align}
leads naturally to a derivative expansion to arbitrary order in the number of derivatives such that the powers of $\hbar$ count the number of derivatives.

Following Ref. \cite{kaynak}, we observe that the resolvent symbol of a bosonic operator such as $-\partial^2 + V(x)$, $V(x)$ being a function of a scalar field $\phi(x)$ with the choice $\hbar$ being set to 1, is given by
\begin{align}
   \sigma\left( \frac{1}{\lambda + A} \right) &= \frac{1}{\left( \lambda + \tilde{A} \right)} - \frac{1}{2 \left( \lambda + \tilde{A} \right)^3} \frac{\partial^2 \tilde{V}}{\partial x^\mu \partial x_\mu} \nonumber \\
   & \quad + \frac{1}{2 \left( \lambda + \tilde{A} \right)^4} \frac{\partial \tilde{V}}{\partial x^\mu} \frac{\partial \tilde{V}}{\partial x_\mu} + \frac{p^\mu p^\nu}{\left( \lambda + \tilde{A} \right)^4} \frac{\partial^2 \tilde{V}}{\partial x^\mu \partial x^\nu} + \cdots \,.
\end{align}
After taking the spectral and the $3$-dimensional momentum integral successively, and then reorganizing the derivative terms by partial integration in Eq.~(\ref{zeta2}), one can observe that the zeta function in ordinary space-time takes the following form,
\begin{align}
   \zeta(s|A) &= \int d^3 x \left[ \frac{\Gamma \left(s-\frac{3}{2}\right) V^{\frac{3}{2}-s}}{8 \pi ^{3/2} \Gamma (s)} -\frac{\Gamma \left(s+\frac{3}{2}\right) V^{-s-\frac{3}{2}}}{96 \pi ^{3/2} \Gamma (s)} \frac{\partial V}{\partial x^\mu} \frac{\partial V}{\partial x_\mu}\right] + \cdots\,,
\end{align}
where the ellipsis stands for the terms containing more derivatives. Plugging minus the derivative of the zeta function above with respect to the complex parameter $s$ at $s=0$  in Eq.~(\ref{deffact}) leads to the desired result for Eq.~(\ref{deffact}) in $2+1$ dimensions. Therefore, the one-loop correction to the Euclidean effective action reads
\begin{align} \label{effact21}
   \Gamma^{(1)} = \frac{1}{2} \int d^3 x \left( -\frac{V^{3/2}}{6 \pi } + \frac{1}{192 \pi} \frac{1}{V^{3/2}} \frac{\partial V}{\partial x^\mu} \frac{\partial V}{\partial x_\mu} \right) + \cdots\,.
\end{align}
This result has a novel feature since each term in it is generally found under different assumptions and limits while using usual derivative expansion methods. In Ref. \cite{ball}, the one loop correction for real scalar fields coupled to a non-homogeneous background field is given as derivative expansion under the assumption that $\phi$ and its derivatives are sufficiently small, and the resulting kinetic term is exactly the one given is in Eq.~(\ref{effact21}). However, it does not have the potential term. On the other hand, exactly the same potential term is obtained in Ref. \cite{justin} in the large potential limit when the field is strong, and derivative terms are relatively small. In Refs. \cite{elizalde,wipf}, this term, whose coefficient is evaluated numerically, is given as a leading strong field limit for constant backgrounds. It is apparent that Eq.~(\ref{effact21}) generates both limits. Since the first term becomes dominant in the strong field limit whereas the second term dominates in the weak field limit. Hence, the result given in Eq.~(\ref{effact21}) provides for an alternative way to obtain the one-loop correction for scalar theories coupled to non-homogeneous background fields as a derivative expansion. Calculations for the terms with four derivatives can be found in Ref. \cite{gusynin1}.

Here, we would like to clarify the point why the potential term is naturally accessible through our method in $3$ dimensions whereas it is not there in the derivative expansion obtained via usual heat kernel method. However, the potential term can be obtained in even dimensions via the latter. We believe that the following facts underlie this difference. First one comes from the fact that the nature of the expansion that we use is different from the usual short-time expansion of the heat kernel, which is briefly explained in the introduction. The second one is related to the existence of the conformal anomaly in even dimensions so as to lead to the logarithmic terms, even if the field and its derivatives are assumed to be small. The emergence of the logarithms will be shown in the next section through the dimensional regularization.

What we would like to do now is to go one step further and to obtain the one-loop correction in generic odd dimensions. First step to be taken is to evaluate the zeta function for the dimensions $d=2n+1$, $n$ being integer, rather than $2+1$ dimensions. This, in practice, does not correspond to the usual zeta function regularization, since the dimension of the space-time is not fixed to any value. Therefore, the result will be a function of the dimension of the space-time. In other words, a new  renormalization prescription should effectively be chosen due to possible divergences that could be encountered for specific values of the dimension at the end. They can be regularized by dimensional regularization in this case due to an analytic dependence of the result on the dimension of the space-time.

It can be shown that the zeta function in $2n+1$ dimensions is given by
\begin{align} \label{13}
   \zeta(s|A) &\simeq \int d^{2n+1} x \left[ \frac{\Gamma \left(s-n-\frac{1}{2}\right) V^{n+\frac{1}{2}-s}}{(4 \pi)^{n+\frac{1}{2}}\Gamma (s)} -\frac{\Gamma \left(s-n+\frac{5}{2}\right) V^{n-\frac{5}{2}-s}}{12(4 \pi)^{n+\frac{1}{2}}\Gamma (s)} \frac{\partial V}{\partial x^\mu} \frac{\partial V}{\partial x_\mu} \right] + O(\partial^4)\,.
\end{align}
Secondly, one has to use the definition given in Eq.~(\ref{deffact}) so as to evaluate the one-loop correction in generic odd dimensions. After straightforward calculations, the desired result reads
\begin{align} \label{14}
   \Gamma^{(1)} &\simeq \frac{1}{2} \int d^{2n+1} x \left[ - \frac{\Gamma \left(-n-\frac{1}{2}\right) V^{n+\frac{1}{2}}}{(4 \pi)^{n+\frac{1}{2}}} + \frac{\Gamma \left(\frac{5}{2}-n\right) V^{n-\frac{5}{2}}}{12 (4 \pi)^{n+\frac{1}{2}}}  \frac{\partial V}{\partial x^\mu} \frac{\partial V}{\partial x_\mu} \right] +  O(\partial^4)\,.
\end{align}
From this result, it can immediately be noticed that for odd dimensions the one-loop correction to the effective action does not possess any divergence. Since the arguments of the gamma functions above are half integers, and even if they become negative, the gamma functions give finite values. It can easily be seen from the result that the one-loop correction to the potential in a non-constant background field is generically governed by $d/2$ fractional power of the potential, mentioned for constant backgrounds in Ref. \cite{wipf}. Moreover, it is obvious that the result is independent of any renormalization scale $\mu$ because for a generic odd dimension the coefficient of the potential term $\Gamma(-n-1/2)$ has a well-defined limit such that $\Gamma(-n-1/2) V^{n+1/2}$ does not generate any logarithm, as a result we do not need to introduce any renormalization scale to make the arguments dimensionless. We would like to emphasize that this is nothing but the manifestation of an important and well-known consequence of the vanishing of the conformal anomaly in odd dimensions from the dimensional regularization point of view. This is, of course, not the case for even dimensions, which will be seen in the next section.
\section{Quantum Effective Action in Even Dimensions} \label{sec3}
On the other hand, it is known from our experiences that divergent expressions occur in even dimensional scalar theories. In order to observe the structure of the divergence for scalar theories at one-loop order, we will evaluate the even dimensional analogs of Eqs.~(\ref{13}) and (\ref{14}). By straightforward calculations, one can, respectively, give the zeta function and the one-loop correction in $2n$ dimensions, $n$ being integer,
\begin{align}
   \zeta(s|A) &\simeq \int d^{2n} x \left[ \frac{\Gamma (s-n) V^{n-s}}{(4 \pi )^{n} \Gamma (s)} -\frac{\Gamma (-n+s+3) V^{n-s-3}}{12 (4 \pi )^{n} \Gamma (s)} \frac{\partial V}{\partial x^\mu} \frac{\partial V}{\partial x_\mu}\right] +  O(\partial^4) \,, \\
   \Gamma^{(1)} &\simeq \frac{1}{2} \int d^{2n} x \left[ -\frac{\Gamma (-n) V^n}{(4 \pi )^{n}} + \frac{\Gamma (3-n) V^{n-3}}{12(4 \pi )^{n}} \frac{\partial V}{\partial x^\mu} \frac{\partial V}{\partial x_\mu}\right] +  O(\partial^4) \,. \label{effeven}
\end{align}
An immediate observation can be made about the kinetic term in Eq.~(\ref{effeven}). It is apparent that a nontrivial wave function renormalization occurs when $2n \geq 6$ at one-loop. This is the case for the $\phi^3_6$ theory, but not the case for the $\phi^4_4$ theory. For the latter one, we have to go beyond the one-loop. It is seen that for all even dimensions, one encounters divergences in the term correcting the classical potential, and as one takes the limit of the parameter $n$ towards physical dimensions in the term $\Gamma(-n) V^n$, it generates logarithmic terms in $\phi(x)$. It is also realized that similar logarithmic terms multiply the kinetic term whenever $2n \geq 6$. In order to observe the differences between the self-interacting scalar theories and the ones coupled to fermionic systems, we recall that in Ref. \cite{kaynak} it is demonstrated that the kinetic term of the scalar sector in the effective action already gains a similar logarithmic multiplier at one-loop in $3+1$ dimensions for the Yukawa theory when the fermionic degrees of freedoms are integrated out. This stems from the fact that a nontrivial wave function renormalization takes place even at that order, which is well-known from the dimensional regularization point of view.

For the sake of completeness, we will demonstrate the transition between two renormalization prescriptions utilized in this paper at one-loop. What is well known from the renormalization theory is that after curing the divergent expression with the assistance of a renormalization prescription, one has to impose some renormalization conditions appropriate to the chosen prescription so as to fix the residual finite parts.

Let's choose the massless $\phi^4_4$ theory. It is a well-known fact that all of the singularities  at one loop can be removed by analytical continuation in zeta function regularization \cite{ball}. Therefore, one can end up with the following effective potential for this theory \cite{ramond},
\begin{align}
V_{eff}(\phi) &= \frac{\lambda \phi^4}{4!} + \frac{\hbar}{64 \pi^2} \frac{\lambda^2 \phi^4}{4} \left[ - \frac{3}{2} + \log \left( \frac{\lambda \phi^2}{2 \mu^2} \right) \right] \,,
\end{align}
where the scale $\mu$ is introduced for dimensional bookkeeping of the logarithm. At the same time, it reflects an occurrence of a renormalization ambiguity by introducing an arbitrary renormalization scale into the theory. Imposing the following condition, corresponding to the coupling constant $\lambda$,
\begin{align} \label{condition}
\lambda &= \frac{d^4 V}{d \phi^4} \bigg|_{\phi=M} \,,
\end{align}
requires,
\begin{align} \label{condition1}
\log \left( \frac{\lambda M^2}{2 \mu^2} \right) &= -\frac{8}{3} \,.
\end{align}

On the other hand, if one chooses the dimensional regularization instead of the zeta function regularization for the same theory by taking the limit $n \rightarrow 2$ in Eq.~(\ref{effeven}) the effective potential, then, takes the following form
\begin{align}
V_{eff}(\phi) &= \frac{\lambda \phi^4}{4!} + \frac{\hbar}{2} \frac{\lambda^2 \phi^4}{4} \left[ -\frac{3}{64} + \frac{1}{16 \pi^2 (d-4)} + \frac{1}{32 \pi^2} \log \left( \frac{\lambda \phi^2}{8 \pi \mu^2} \right) \right] \,.
\end{align}
After minimal substraction, leading to the redefinition of the coupling constant, by which the pole is absorbed, the condition in Eq.~(\ref{condition}) is ready to be imposed, and it reads
\begin{align} \label{condition2}
\log \left( \frac{\lambda M^2 e^\gamma}{8 \pi \mu^2} \right) &= -\frac{8}{3} \,.
\end{align}
If the conditions in Eqs.~(\ref{condition1}) and (\ref{condition2}) are plugged into the corresponding effective potentials, the same effective potential is found, which is made physical by the renormalization theory,
\begin{align}
V_{eff}(\phi) &= \frac{\lambda \phi^4}{4!} + \frac{\hbar}{64 \pi^2} \frac{\lambda^2 \phi^4}{4} \left[ - \frac{25}{6} + \log \left( \frac{\phi^2}{M^2} \right) \right]\,,
\end{align}
which is the well-known result for the $\phi^4_4$ theory \cite{coleman}. Hence, the renormalization condition related to the coupling constant yields a different equality in each prescription. They resolve the ambiguity resulted from the introduction of a generic renormalization scale, and fix the finite part in each case in such a way that we obtain the same result at one-loop.
\section{The $O(N)$-Invariant Nonlinear $\sigma$-Model in $d=2+1$}\label{sec4}
In this section, we will evaluate the beta function of the $O(N)$-invariant nonlinear $\sigma$-model in $2+1$ dimensions through the effective action for the auxiliary field at leading order in the $1/N$ expansion without using $\varepsilon$-expansion. As in Refs. \cite{polyakov,warr}, the Euclidean partition function of the $O(N)$ nonlinear $\sigma$-model in $d=3$ can be given by
\begin{align}
\mathcal{Z} &= \int \mathscr{D} [\phi] \mathscr{D} [\sigma] \exp \left\{ - \int d^3 x \left[ \frac{1}{2} \phi^i \left( - \partial^2 + \sigma \right) \phi_i   - \frac{1}{2 g_0} \sigma \right] \right\} \,,
\end{align}
where the auxiliary field $\sigma(x)$ is a Lagrange multiplier, and it is introduced to impose the constraint $\phi^i(x) \phi_i(x)=1$ in the classical limit. Moreover, it allows us to integrate over the field $\phi_i$ so that the partition function  reads
\begin{align}\label{partition}
\mathcal{Z} &= \int \mathscr{D} [\sigma] \exp \, S_{eff} ( \sigma) \,,
\end{align}
with
\begin{align} \label{effact}
S_{eff} &=   \frac{1}{2 g_0} \int d^3 x \, \sigma(x) - \frac{N}{2} \mathrm{Tr} \log \left( - \partial^2 + \sigma \right) \,.
\end{align}
Since the limit $N \rightarrow \infty$ with $N g_0$ held fixed will be taken, the exponent in Eq.~(\ref{partition}) is of order $N$. In this respect, one can thus evaluate the functional integral via the saddle point approximation. The value of the field $\sigma(x)$ that minimizes exponent will dominate the path integral in that limit. Before the evaluation of the path integral, we should impose the renormalization condition. Since there is not any nontrivial wave-function renormalization, the first term in Eq.~(\ref{effact21}) is sufficient for our purpose. Therefore, Eqs.~(\ref{effact21}) and~(\ref{effact}), disregarding the derivative term gives
\begin{align}
\frac{\sigma}{2 g_0} + \frac{N \sigma^{3/2}}{12 \pi} \,,
\end{align}
which is equivalent to evaluating the determinant in constant background. If one takes the auxilary field constant, and imposes the renormalization condition corresponding to the coupling constant in order to fix the finite residual part, one can obtain
\begin{align}\label{finite}
\frac{d}{d \sigma} \left[ \frac{\sigma}{2 g_0} + \frac{N \sigma^{3/2}}{12 \pi} \right]_{\sigma=\mu^2} &= \frac{\mu}{2 g(\mu)} \,.
\end{align}
Here, the redefinition of the coupling constant $1/g(\mu)$ to $\mu/g(\mu)$ makes the coupling constant dimensionless. The condition~(\ref{finite}) leads to a finite shift in the coupling constant, and it reads
\begin{align}\label{coupling}
\frac{1}{g_0} &= \frac{\mu}{ g(\mu)} - \frac{N \mu}{4 \pi} \,,
\end{align}
where the second term on the right-hand side implies that linear divergence appears if the theory is regularized by the cutoff regularization rather than the zeta-function regularization. After replacing the coupling constant by its shifted version, the effective action can be given by
\begin{align}
S_{eff}(\sigma) &= \frac{\mu}{2 g(\mu)} \int d^3 x \,\sigma(x) - \frac{N \mu}{8 \pi} \int d^3 x\, \sigma(x) - \frac{N}{2} \mathrm{Tr} \log \left( - \partial^2 + \sigma \right) \,.
\end{align}
The saddle point equation is obtained by extremizing the action with respect to the auxiliary field $\sigma(x)$ set to a constant, and it reads in a dimensionless fashion
\begin{align}
\frac{d}{d \sigma} \bigg|_{\sigma=m^2} \left[  \frac{1}{2 g(\mu)} \left (\frac{\sigma}{\mu^2} \right) - \frac{N}{8 \pi} \left( \frac{\sigma}{\mu^2} \right) + \frac{N}{12 \pi} \left( \frac{\sigma}{\mu^2} \right)^{3/2} \right] &= 0 
\end{align}
The extremum is at the point
\begin{align}\label{mass}
\sigma &= m^2 = \frac{\mu ^2 (g N-4 \pi )^2}{g^2 N^2} \,.
\end{align}
Since the minimum $m^2$ should be renormalization group invariant, it should obey the equation,
\begin{align}
\left( \mu \frac{\partial}{\partial \mu} + \beta(g) \frac{\partial}{\partial g} \right) m^2 &= 0 \,.
\end{align}
This allow us to compute the beta function of the theory. As a result, the $\beta$ function can be given by
\begin{align}\label{beta}
\beta(g) &= g - \frac{N g^2}{4 \pi} \,,
\end{align}
in accordance with the result in Ref. \cite{justin} as $\varepsilon \rightarrow 1$ in $\varepsilon$-expansion. It is apparent that holding $N g$ kept fixed as $N \rightarrow \infty$ either in Eq.~(\ref{coupling}) or in Eq.~(\ref{beta}) sets the critical value of the coupling constant $g_c = 4 \pi$ at leading order in the $1/N$ expansion, as given in Ref. \cite{arafeva} for the nonlinear chiral field. In that limit, the  mass $m$ goes to zero at $g=g_c$, which can be easily seen from Eq.~(\ref{mass}). The theory, thus, becomes critical at this nontrivial fixed point, which is UV stable.
\section{Acknowledgements}
The author would like to thank Prof. O. Teoman Turgut for reading the manuscript, and Prof. V. Gusynin for bringing our attention to Ref. \cite{gusynin1}, where similar results for the effective action were presented.

\end{document}